\documentclass[twocolumn,showpacs,preprintnumbers,amsmath,amssymb]{revtex4}

\usepackage{graphicx}
\usepackage{dcolumn}
\usepackage{bm}

\begin{document}

\title{First Excited \vspace{1pt} Band of a Spinor Bose-Einstein
Condensate}

\author{C.G. Bao\vspace{1pt} and Z.B. Li}

\affiliation{The State Key Laboratory of Optoelectronic Materials
and Technologies \\ Department of Physics, Zhongshan University,
Guangzhou, 510275, P.R. China}

\vspace{1pt}

\begin{abstract} The analytical expression of the fractional parentage
coefficients for the total spin-states of a spinor N-boson system
has been derived. Thereby an S-conserved theory for the spinor
Bose-Einstein condensation has been proposed. A set of equations
has been established to describe the first excited band of the
condensates. Numerical solution for $^{23}$Na has been given as an
example.
\end{abstract}

\maketitle

\vspace{1pt}

In recent years, accompanying the experimental realization of the
spinor Bose-Einstein condensation
(BEC)\cite{ref1,ref2,ref3,ref4,ref5}, corresponding theories based
on the mean field theory have been developed\cite{ref6,ref7, ref8,
ref9, ref10, ref11, ref12, ref13}. The total spin S together with
its Z-component are actually conserved in the spinor condensates.
Furthermore, the interactions are in general spin-dependent.
Therefore, the strength of the mean field $g$ that acts on the
macroscopically occupied quantum states might depend on S. In
previous theories, the S-dependence of the strength $g$ has not
yet been perfectly clarified. A primary attempt along this line
was proposed in Ref. \cite{ref13b}  where the total spin-states
have been introduced and the S-dependence of $g$ has been derived
analytically.

The present paper is a continuation of the Ref. \cite{ref13b}. It
is noted that the thermodynamical property of condensates at low
temperature depends on their low-lying spectra, this is an
interesting topic scarcely touched. The first aim of the paper is
to generalize the study of Ref. \cite{ref13b} from the ground band
(GB) to excited bands. Rigorous total spin-states with good
quantum numbers, $S$ and $S_{Z}$, and with a specific permutation
symmetry (see below)  will be introduced, a set of equations is
derived to describe, as a first step, the first excited band (FEB)
of the spinor BEC.

When the total spin states are adopted, how to achieve the matrix
elements among them is crucial. The second aim of this paper is to
develop a general tool, namely the fractional parentage
coefficients, to facilitate the theoretical treatment of spinor
N-boson systems. The case of $^{23}$Na as an example will be
studied numerically. Before going ahead, we need some knowledge
from few-boson systems.

Let the interaction between a pair of spin-one bosons be $%
U_{ij}=\delta (\mathbf{r}_{i}\mathbf{-r}_{j})O_{ij}$ , $%
O_{ij}=g_{m}P_{m}+g_{q}P_{q}=c_{0}+c_{2}\mathbf{F}_{i}\cdot %
\mathbf{F}_{j}$\ . \ Where $P_{m}$ and $P_{q}$ \ are projectors to
the two-body spin-states with spin $S_{ij}=0$ and 2\ (spin $1$ is
irrelevant since in this case both the spin wavefunction and the
spatial wavefunction must be antisymmetric that prevents a point
interaction), \ respectively,\ \ $g_{m}$ and $g_{q}$\ are the
associated strengths, \ $\mathbf{F}_{i}$ is the spin operator of particle $i$%
, $c_{0}=(g_{m}+2g_{q})/3$\ , and \ $c_{2}=(g_{q}-g_{m})/3$. \ Let the $%
N $ bosons (with mass $m$) be confined by a harmonic trap with frequency $%
\omega $. \ When $\hbar \omega $ and $\sqrt{\frac{\hbar }{m\omega }}$ are
used as units of energy and length, respectively, the Hamiltonian reads
\begin{equation}
H=\frac{1}{2}\sum_{i}(-\nabla _{i}^{2}+r_{i}^{2})+\sum_{i<j}U_{ij}
\label{Ham}
\end{equation}

Using the method as given in \cite{ref14}, the Hamiltonians for $N=$ $%
3$ and $4$ \ two-dimensional\ systems have been diagonalized. Both
the orbital angular momentum $L$\ and $S$ are good quantum
numbers, therefore the eigenenergies and eigenstates can be
denoted as $E_{LS}$\ and $\Psi _{LS} $\ (there is actually a
series of states having the same $L$ and \ $S,$ but only the
lowest of them is concerned)\ . \ It was found that, the levels
are grouped  as shown in Fig.1. In the lowest (first) group all
bosons remain mostly in the lowest harmonic oscillator (h.o.)
shell, in the second group one of the bosons is mainly in the
first excited h.o. shell, etc. .\ All the levels in the first
group have $L=0$\ , they would be degenerate if the interaction
were spin-independent (i.e., $g_{q}=g_{m}$\ or \ $c_{2}=0$\ ),\
otherwise they would split into a band, the ground band (GB).  \
All levels in the second group have \ $L=1$\ , \ they would be
degenerate into two levels if \ $g_{q}=g_{m}$ \ , \ and each would
split into a band if \ $g_{q}\neq g_{m}$ \ . \ Thus, the two
lowest group contain totally three bands. In particular, we found
that the third band is exactly a shift as a whole of the ground
band by the unit \ $\hbar \omega $ \ , \ therefore it is simply
the c.m. excited states of the ground band. We shall concentrate
on the second band, namely the FEB.

Since the total wave function must be totally symmetric with
respect to particle permutation, it can be in general written as
\begin{equation}
\Psi _{LS}=\sum_{k,\lambda j}f_{LSk,\lambda j}^{\;}\vartheta
_{S,k}^{\lambda j} \label{eq2}
\end{equation}
where \ $\vartheta _{S,k}^{\lambda j}$ \ is the total spin-state
with the good quantum numbers $S$ \ ,\ $S_{Z}$ \ ,\ and it is also
the \ $j-th$ \ basis function of the \ $\lambda$ \ representation
of the \ $N$-body permutation group. The index \ $k$ \ refers to
the multiplicity, i.e., more than one set of total spin states
might have the same $S$ and $\lambda $\ .\ \cite{ref15} \ $f_{LSk,
\lambda j}$ \ is a spatial function belonging to the same
representation $\lambda $ \ . \ The summation over $j$\ is necessary to assure $%
\Psi _{LS}$ to be totally symmetric. The summation over $\lambda $\ and $%
k $\ is necessary because that they are not conserved under the
interaction.

For a given state, from (\ref{eq2}) we can define the weight of a
representation as
\begin{equation}
W_{LS,\lambda }=\sum_{k,j}<f_{LSk,\lambda j}^{\;}|f_{LSk,\lambda
j}^{\;}> \label{eq3}
\end{equation}
When $ g_{q}=g_{m} $\   the permutation symmetry is pure and
therefore\ $W_{LS,\lambda }=1$ or $0$ \ . \ When $g_{q} \neq
g_{m}$\ , \ the mixing of $\lambda $ appears, but remains very
weak. It was found that, for both $N=3$ and $4$, the GB is
dominated by $\lambda=\{N\}$ , while the FEB by
$\lambda=\{N-1,1\}$. Even if the spin-dependent interaction is
very strong ($g_{q}/g_{m}=0.6$ and $g_{m}+g_{q}=3)$ , the GB still
has $W_{0S,\{N\}}\geq 0.996$ and the FEB has $W_{1S,\{N-1,1\}}\geq
0.998$ , thus the mixing is still weak. Since the mixing is so
slight, $\lambda$ is nearly a good quantum number. Therefore, the
labels $(L, S, \lambda)$ can be used to label the states as marked
in Fig.1.

There are four points noticeable.

(i) The zero-range interaction between two bosons would be zero if
their spatial wave function is antisymmetric (because in this case
the two bosons do not overlap). \ Since the interaction is
repulsive, the representation $\lambda $ \ with  more
antisymmetric pairs is more advantageous to binding because the
repulsion can be thereby reduced.

(ii) When one and only one boson has been excited to the second
h.o. shell, the spatial permutation symmetry is restricted to
$\lambda =\{N\} $ and $\{N-1,1\},$ and this explains why the
second group contains two bands as shown in Fig.1. \ Where the
band with \{$N-1,1$\} is lower due to the reason stated in point
(i). \ Thus the FEB would in general have $\lambda =\{N-1,1\}$ \ .

(iii) The energy would also depend on \ $S$ \ due to the
spin-dependent interaction. Thus, the levels with the same
$\lambda $ \ would further split into a band.

(iv) When $g_{q}>g_{m}$\ , the pair coupled to \ $S_{ij}=2$ \ is
more repulsive than the pair coupled to 0 \ . \ Evidently, the
many-body state with a larger \ $S$ \ would have more pairs
coupled to 2. \ Therefore, in a band, the state with a larger $S$\
is higher. On the contrary, when \ $g_{q}<g_{m} $\ , \ the state
with a larger $S$ is lower as shown in Fig.1.

To generalize the above study to the case with a very large $N$,
we need the knowledge of the total spin-states. For the most
important case of $\lambda=\{N\}$, there is an unique
$\vartheta^{\{N\}j}_{S,k}$ if $N-S$ is even, or there is none
otherwise\cite{ref17b, ref13b}. Therefore both the labels $j$ and
$k$ are not necessary, and a simpler notation
$\vartheta^{\{N\}}_{S}\equiv\vartheta^{\{N\}j}_{S,k}$ can be used.

Let us start from a 2-boson system with %
$\vartheta _{S}^{\{2\}}(12)=(\chi (1)\chi (2))_{S}$\ , \ where $%
\chi (i)$ is the single boson spin-state, $S=0$ or 2. \ For \
$N=3$\ , \ the requirement that \ $N-S$\ is even leads to \ $S=3$\
and 1. \ They were given in \cite{ref14} as
\begin{equation}
\vartheta _{3}^{\{3\}}=\{(\chi (1)\chi (2))_{2}\chi
(3)\}_{3}=\{\vartheta _{2}^{\{2\}}\chi (3)\}_{3} \label{th3}
\end{equation}
\begin{equation}
\vartheta _{1}^{\{3\}}=\frac{2}{3}\{\vartheta _{2}^{\{2\}}\chi
(3)\}_{1}+\frac{\sqrt{5}}{3}\{\vartheta _{0}^{\{2\}}\chi
(3)\}_{1}\label{th1}
\end{equation}

Starting from (\ref{th3}) and (\ref{th1}), let us assume that all
\ $\vartheta _{S^{\prime }}^{\{N-1\}}$ \ of the \ ($N-1$)-boson
system have been known. Then, taking the rule of angular momentum
coupling and the rule of outer product into account, the \
$\vartheta _{S}^{\{N\}}$ \ of the $N$-boson system can be uniquely
expanded as
\begin{equation}
\vartheta _{S}^{\{N\}}=a_{S}^{\{N\}}\{\vartheta
_{S+1}^{\{N-1\}}\chi (N)\}_{S}+b_{S}^{\{N\}}\{\vartheta
_{S-1}^{\{N-1\}}\chi (N)\}_{S} \label{thS}
\end{equation}
where\ $a_{S^{\prime }}^{\{N\}}$ \ and \ $b_{S^{\prime }}^{\{N\}}$
\ are the fractional parentage coefficients (FPC) to be
determined, however they are zero if $N-S$ is odd.

From (\ref{thS}), a similar formula can be written for \
$\vartheta _{S\pm 1}^{\{N-1\}}$\ . \ Therefore, \ $\vartheta
_{S}^{\{N\}}$ \ can be expanded once more as
\begin{eqnarray}
\vartheta _{S}^{\{N\}}&=&a_{S}^{\{N\}}a_{S+1}^{\{N-1\}}\{\lbrack
\vartheta _{S+2}^{\{N-2\}}\chi (N-1)\rbrack _{S+1}\chi (N)\}_{S}
\nonumber \\ &+& a_{S}^{\{N\}}b_{S+1}^{\{N-1\}}\{\lbrack \vartheta
_{S}^{\{N-2\}}\chi (N-1)\rbrack _{S+1}\chi (N)\}_{S} \nonumber \\
&+& b_{S}^{\{N\}}a_{S-1}^{\{N-1\}}\{\lbrack \vartheta
_{S}^{\{N-2\}}\chi (N-1)\rbrack _{S-1}\chi (N)\}_{S} \nonumber \\
&+& b_{S}^{\{N\}}b_{S-1}^{\{N-1\}}\{\lbrack \vartheta
_{S-2}^{\{N-2\}}\chi (N-1)\rbrack _{S-1}\chi (N)\}_{S} \nonumber \\
\label{thSN}
\end{eqnarray}

On the other hand, since the $\vartheta _{S}^{\{N\}}$ is totally
symmetric as required, the right hand side of (\ref{thSN}) must be
invariant against the interchange of the \ N-th and (N-1)-th\
bosons. This leads to a set of three homogeneous equations. It
turns out that they are linearly dependent, one of them reads
\begin{eqnarray}
~& &\lbrack\stackrel{\wedge }{W_{S}}(S+1,S+1)-1\rbrack
a_{S}^{\{N\}}b_{S+1}^{\{N-1\}} \nonumber \\
&+&\stackrel{\wedge }{W_{S}}(S-1,S+1)b_{S}^{\{N%
\}}a_{S-1}^{\{N-1\}}=0 \label{WS1}
\end{eqnarray}
where
\begin{equation}
\stackrel{\wedge }{W_{S}}(T,T^{\prime })=\sqrt{(2T+1)(2T^{\prime }+1)%
}W(1SS1;TT^{\prime }) \label{WST}
\end{equation}
and \ $W(1SS1;TT^{\prime })$\ is the well known Racah
coefficients. From (\ref{WS1}) we obtain
\begin{equation}
a_{S}^{\{N\}}=\gamma^{\{N\}}_{S} \frac{\sqrt{2S-1}}{S\sqrt{2S+1}}%
/b_{S+1}^{\{N-1\}} \label{aNS}
\end{equation}
\begin{equation}
b_{S}^{\{N\}}=\gamma^{\{N\}}_{S} \frac{\sqrt{2S+3}}{(S+1)\sqrt{2S+1}}%
/a_{S-1}^{\{N-1\}} \label{bNS}
\end{equation}
if $N-S$ is even, where the constant\ $\gamma _{S}^{\{N\}}$\ is introduced to assure \ $%
(a_{S}^{\{N\}})^{2}+(b_{S}^{\{N\}})^{2}=1$ \ ,\ so that $\vartheta
_{S}^{\{N\}}$ \ is normalized. \ Eqs.(\ref{aNS}) and (\ref{bNS})
together with the condition of normalization lead to a unique
solution, thus we obtain the analytical form of FPC as
\begin{equation}
a_{S}^{\{N\}}=\lbrack (1+(-1)^{N-S})(N-S)(S+1)/(2N(2S+1))\rbrack
^{1/2}       \label{aNSexpr}
\end{equation}
\begin{equation}
b_{S}^{\{N\}}=\lbrack (1+(-1)^{N-S})\;S\;(N+S+1)/(2N(2S+1))\rbrack
^{1/2}\label{bNSexpr}
\end{equation}
and
\begin{equation}
\gamma_{S}^{\{N\}}=S(S+1)\lbrack
\frac{(N-S)(N+S+1)}{N(N-1)(2S+3)(2S-1)}\rbrack ^{1/2}
\label{gNSexpr}
\end{equation}

The FPC are very useful because related matrix elements among the
total spin-states can be thereby calculated.

Now let us study the set of total spin-states with the \
$\{N-1,1\}$\ symmetry. It has been proved that the multiplicity of
this set is one \cite{ref15}, therefore they can be labelled in a
simpler way as \ $\Theta _{S}^{i}$ $\equiv $ $\vartheta
_{S,k}^{\{N-1,1\}i}$\ . Since the dimension of $\{N-1,1\}$\
representation is $N-1$\ , $i=1,2, \cdots, N-1$\ .\  Making use of
the FPC, they can be expanded as
\begin{equation}
\Theta _{S}^{i}=b_{S}^{\{N\}}\{\vartheta _{S+1}^{\{N-1\}}(\stackrel{%
\times }{i})\chi (i)\}_{S}-a_{S}^{\{N\}}\{\vartheta _{S-1}^{\{N-1\}}(%
\stackrel{\times }{i})\chi (i)\}_{S} \label{ThSeven}
\end{equation}
if \ $N-S$\ is even, and
\begin{equation}
\Theta _{S}^{i}=\{\vartheta _{S}^{\{N-1\}}(\stackrel{%
\times }{i})\chi (i)\}_{S} \label{ThSodd}
\end{equation}
if $N-S$ \ is odd, where the notation \ $\stackrel{\times }{i}$ \
represents all the bosons except the $i-th$\ . \ From the rule of
outer product, (\ref{ThSeven}) and (\ref{ThSodd}) \ will contain
only the symmetries \ $\{N\}$ \ and \ $\{N-1,1\}$ \ . \ However,
when \ $N-S$ \ is even, (\ref{ThSeven}) is orthogonal to the
unique totally symmetric spin-state\ $\vartheta _{S}^{\{N\}}$\ . \
Hence, it has the pure\ $\{N-1,1\}$ \ symmetry. \ When\ $N-S$\ is
odd, it has been stated above that\ $\vartheta
_{S}^{\{N\}}=0$\ , therefore\ (\ref{ThSodd}) \ has also the pure $%
\{N-1,1\}$ symmetry. It is obvious from (\ref{aNSexpr}) and
(\ref{bNSexpr}) that\ $\Theta _{S}^{i}$ \ does not exist if $S=0$\
or $N$\ .\ Thus the $S$ of \ $\Theta _{S}^{i}$ \ is ranged from 1
to $N-1$\ , \ while the index $i$\ of \ $\Theta _{S}^{i}$ \ is
ranged from 1 to $N$ . This set of $N$ states are not mutually
orthogonal but linearly dependent. They satisfy $\sum_{i}\Theta
_{S}^{i}=0$. \ This arises because the summation over $i$\ leads
to a symmetrization, a state with \ $\lambda \neq \{N\}$ \ will
not survive after the symmetrization.

Based on the knowledge of $\Theta _{S}^{i}$\ , we can begin to
study the FEB (three-dimensional) of realistic cases. Let the
normalized single-particle state of
condensation be denoted as $\varphi _{S}^{a}=\frac{1}{\sqrt{4\pi }\;r}%
u_{S}^{a}(r)$\ ,\ while the excited state be denoted as $\varphi _{S}^{b}=%
\frac{1}{\;r}u_{S}^{b}(r)Y_{lm}(\theta, \phi)$\ with \ $l=1$\ . \
As a basic assumption we define the spatial wave function as
\begin{equation}
F_{S,i}=\varphi _{S}^{b}(i)\Pi_{j(\neq i)} \varphi _{S}^{a}(j)
\label{eq11}
\end{equation}
Then, the total wave function is written as
\begin{equation}
\Psi _{S}=\sum_{i}\;F_{S,i}\;\Theta _{S}^{i} \label{eq12}
\end{equation}
Where the summation over $i$ leads to a symmetrization as
required, and the $\{N-1,1\}$-representation has been introduced.
The expansion in (18) is different from (2), but both are
eigenstates of the total spin and symmetrical with respect to
particle permutation, thus both are applicable.

Inserting (\ref{eq12}) into the Schr\"{o}dinger equation, using
the standard variational procedure, the equations for $\varphi _{S}^{a}$ and $%
\varphi _{S}^{b}$ can be deduced as
\begin{equation}
\lbrack h_{0}+g_{S}^{b}\frac{|u_{S}^{b}|^{2}}{4\pi r^{2}}+(N-2)g_{S}^{a}%
\frac{|u_{S}^{a}|^{2}}{4\pi r^{2}}\rbrack \;u_{S}^{a}=\varepsilon
_{a}u_{S}^{a}  \label{cgp1}
\end{equation}
\begin{equation}
\lbrack h_{1}+(N-1)g_{S}^{b}\frac{|u_{S}^{a}|^{2}}{4\pi r^{2}}\rbrack
\;u_{S}^{b}=\varepsilon _{b}u_{S}^{b}  \label{cgp2}
\end{equation}
where
\begin{equation}
h_{0} =\frac{1}{2}\lbrack -\frac{d^{2}}{dr^{2}}+r^{2}\rbrack
\label{h0}
\end{equation}
\begin{equation}
h_{1} =\frac{1}{2}\lbrack -\frac{d^{2}}{dr^{2}}+\frac{2}{r^{2}}%
+r^{2}\rbrack  \label{h1}
\end{equation}
\begin{equation}
g_{S}^{b}=\langle \Theta _{S}^{i}|O_{ij}\;|\Theta _{S}^{i}+\Theta
_{S}^{j}\rangle  \label{defgSb}
\end{equation}
and
\begin{equation}
g_{S}^{a}=\langle \Theta _{S}^{i}|O_{jk}\;|\Theta _{S}^{i}\rangle
\label{defgSc}
\end{equation}

The energies of the states in the FEB are given by
\begin{eqnarray}
E_{S}^{(e)} &=&\varepsilon _{b}+(N-1)\varepsilon _{a}  \nonumber \\
&-&\frac{(N-1)g_{S}^{b}}{4\pi }\int_{0}^{\infty }\frac{dr}{r^{2}}%
|u_{S}^{a}|^{2}|u_{S}^{b}|^{2}  \nonumber \\
&-&\frac{(N-1)(N-2)g_{S}^{a}}{8\pi }\int_{0}^{\infty }\frac{dr}{r^{2}}%
|u_{S}^{a}|^{4}  \label{Ese}
\end{eqnarray}
where $S$ can be even or odd, and is ranged from $1$ to $N-1$.

Using the expansions (\ref{ThSeven}) and (\ref{ThSodd}) twice (one
is for the \ $N-1$\ body system), with the FPC given by
(\ref{aNSexpr}) and (\ref{bNSexpr}), the matrix elements among \
$\Theta _{S}^{i}$ \ can be obtained as
\begin{equation}
g_{S}^{b}=\frac{N-2}{N-1}(c_{0}+c_{2})-(1+(-1)^{N-S})\frac{S(S+1)}{N\;(N-1)}c_{2}
\label{gSbeven}
\end{equation}

\begin{eqnarray}
g_{S}^{a}&=&c_{0}+ c_{2}\frac{1\;}{(N-1)(N-2)} \cdot \nonumber \\
& &\lbrack \frac{N+1+(-1)^{N-S}}{N}%
S(S+1)-2(N-1)\rbrack  \label{gSceven}
\end{eqnarray}

In the FEB, all the members with odd $N-S$ do not have their
partners in the ground band, they are the lowest states with the
given $S$. \ If the system is cooled from higher to zero
temperature, and if no spin-relaxation mechanism is available, the
probability of staying in the FEB is nearly half. \ \
Incidentally, when $N$ is large, since $g_{S}^{a}$ is very close
to the $g_{S}=c_{0}+(S(S+1)-2N)/(N(N-1))\;c_{2}$\ \ of the ground band$%
^{14}$ , thus $\varphi_{S}^{a}$ is very close to that of the
ground band if $N-S$\ is even.

As an application we calculated the FEB of $^{23}$Na\ atoms
trapped in a harmonic well with $\hbar \omega =100Hz$\ and with
$g_{m}=6.351\times 10^{-4} \sqrt{\omega }$ , $g_{q}/g_{m}=1.10$ \
.  The normalized radial wave functions $R^{a}(r)=u_{S}^{a}(r)/r$
and $R^{b}(r)=u_{S}^{b}(r)/r$ are given in Fig.2\ (a) and (b)
respectively.

Let $E_{S}^{(g)}$ denote the energies of the GB ($E_{0}^{(g)}$ is
the ground state energy ). The upper and lower bounds of the FEB,
i.e., $E_{N-1}^{(e)}$ and $E_{1}^{(e)}$ relative to $E_{0}^{(g)}$
are plotted in Fig.3. This figure demonstrates how the FEB and GB
split. In particular when $N$ gets larger, the lower bound of FEB
would cross the upper of the GB, it implies an overlap of these
two bands.

In summary, we have derived a set of equations to describe the
first excited band, this might lead to a deeper understanding of
the condensates at very low temperature. Furthermore, we have
achieved the analytical form of the fractional parentage
coefficients. This will greatly facilitate the calculations
dealing with the total spin-states, and will thereby promote the
applicability of the S-conserved theory for various spinor
many-body systems.

Acknowledgment; We appreciate the support by NSFC under the grants 90103028
and \ 90306016\ .

\clearpage

\begin{figure}
\includegraphics{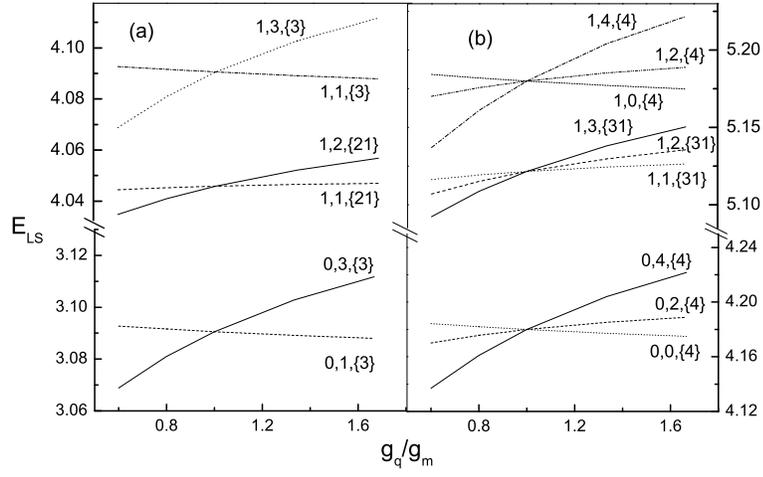}
\caption{\label{fig:1} The evolution of $E_{LS} $\ \ (in $\hbar \omega $\ ) with $%
g_{q}/g_{m} $ in 3 -boson (a) and 4-boson (b) systems. \
($L,S,\lambda $) are marked by the levels. }
\end{figure}

\begin{figure}
\includegraphics{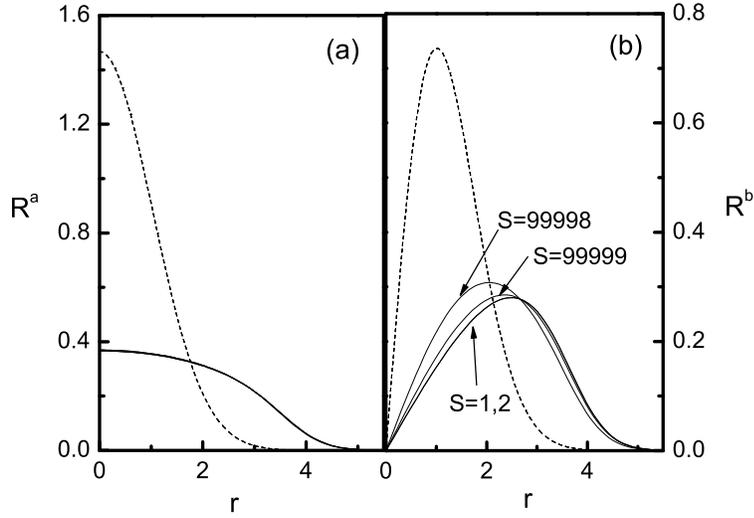}
\caption{\label{fig:2} The radial wave functions, $R^{a}$ (in (a)
) and $R^{b}$ ( in (b) ), as functions of $r$. The dashed curves
are for $N=100$, while the solid ones are for $N=10^{5}$\ . In
each case the curves for $S=N-1, N-2, 2$, and $1$ are plotted.
However, they can not be distinguished except the case of $R^{b}$
with a large $N$, where $S$ are marked by the curves. In (b), when
$S$ lies in between and $N-S$ is even (odd), the associated curves
are distributed between the curves with $S=2$ and $99998$ ($1$ and
$99999$ ). Obviously, when $N-S$ is odd, the S-dependence is less
explicit, and the $R^{b}$ with $S=N-2$ has a smallest size. }
\end{figure}

\begin{figure}
\includegraphics{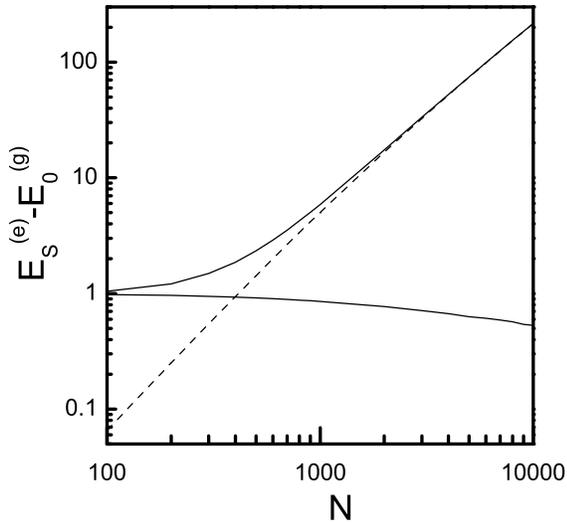}
\caption{\label{fig:3} $E_{S}^{(e)}-E_{0}^{(g)}$ against $N$. The
upper (lower) solid line has $S=N-1$ ($1$). The dashed line is for
$E_{N}^{(g)}-E_{0}^{(g)}$, namely the split of the GB.}
\end{figure}

\end{document}